\documentclass[twocolumn]{aa}
\usepackage{graphicx}
\usepackage{txfonts}
\usepackage[authoryear]{natbib}
\bibpunct{(}{)}{;}{a}{}{,}
\newcommand{\src}  {SWIFT J1626.6--5156}

\def\simless{\mathbin{\lower 3pt\hbox
     {$\rlap{\raise 5pt\hbox{$\char'074$}}\mathchar"7218$}}}   %< or of order
\def\simmore{\mathbin{\lower 3pt\hbox
     {$\rlap{\raise 5pt\hbox{$\char'076$}}\mathchar"7218$}}}   %> or of order

\begin{document}

   \title{Bright flares from the X-ray pulsar SWIFT J1626.6--5156} 

   \subtitle{}

   \author{	
   	P. Reig\inst{1,2}
       \and 
         T. Belloni\inst{3}
	\and 
	G.L. Israel\inst{4}
	\and
	S. Campana\inst{3}
	\and
	N. Gehrels\inst{5}
	\and
	J. Homan\inst{6}
	}

\authorrunning{Reig et al.}
\titlerunning{SWIFT J1626.6--5156}

   \offprints{pau@physics.uoc.gr}

   \institute{
   IESL, Foundation for Research and Technology, 71110, Heraklion, Greece
	\and
   Physics Department, University of Crete, 71003, Heraklion, Greece
	\email{pau@physics.uoc.gr}
	\and
   INAF - Osservatorio Astronomico di Brera, Via E. Bianchi 46, 
   I-23807 Merate (LC), Italy
   	\email{tomaso.belloni@brera.inaf.it, sergio.campana@brera.inaf.it}
	\and
   INAF - Osservatorio Astronomico di Roma, Via Frascati 33, 
   I-00040 Monteporzio Catone, Roma, Italy.
	\email{gianluca@mporzio.astro.it}
	\and
	NASA Goddard Space Flight Center, Greenbelt, MD 20771, USA
	\email{Neil.Gehrels@gsfc.nasa.gov}
	\and
	MIT Kavli Institute for Astrophysics and Space Research, 
	70 Vassar Street, Cambridge, MA 02139, USA
	\email{jeroen@space.mit.edu}
   }

   \date{Received ; accepted}

\abstract
%context
{%\src\ was discovered by SWIFT on December 18, 2005. Besides pulsations,
%the X-ray emission exhibits short-lived flares.
}
%aims
{We have performed a timing and spectral analysis of the X-ray pulsar \src\
during a major X-ray outburst in order to unveil its nature and investigate 
its flaring activity. 
}
%methods
{Epoch- and pulse-folding techniques were used to derive the spin period. 
Time-average and pulse-phase spectroscopy were employed to study the 
spectral variability in the flare and out-of-flare states and energy 
variations with pulse phase. Power spectra were obtained to investigate the 
periodic and aperiodic variability.
}
%results
{Two large flares, with a duration of $\sim$450 seconds were observed on 24
and 25 December 2005. During the flares, the X-ray intensity increased by a
factor of 3.5, while the peak-to-peak pulsed amplitude increased from 45\%
to 70\%. A third, smaller flare
of duration $\sim$180 s was observed on 27 December 2005. The flares seen in 
\src\ constitute the shortest events of this kind ever reported in a 
high-mass X-ray binary. In addition to the flaring activity, strong X-ray 
pulsations with $P_{\rm spin}=15.3714\pm0.0003$ s characterise the X-ray 
emission in \src. After the major outburst, the light curve exhibits 
strong long-term variations modulated with a 45-day period. We relate this 
modulation to the orbital period of the system or to a harmonic.
Power density spectra show, in addition to the harmonic components of the
pulsation, strong band-limited noise with an integrated 0.01-100 Hz
fractional rms of around 40\% that increased to 64\% during the flares. A
weak QPO (fractional rms 4.7\%) with characteristic frequency of 1 Hz was 
detected in the non-flare emission. 
The timing (short X-ray pulsations, long orbital period) and spectral (power-law with cut off energy 
and neutral iron line) properties of \src\ are characteristic of Be/X-ray
binaries. 
%In general, the pre-flare pulse shape
%is consistent with an average template, the post-flare profile with a
%de-enhanced (negative Gaussian) template and the flare pulse profile with
%the same average template modified by one or two Gaussians.
}
%conclusions
%{
%The timing (X-ray pulsations) and spectral (power-law with cut off energy 
%and neutral iron line) properties of \src\ are characteristic of Be/X-ray
%binaries. It is the first time that such short-lived flare events are seen 
%in a Be/X-ray binary. We speculate that the flares might be the result
%of accretion from a clumpy wind.
%} 


   \maketitle

\section{Introduction}

\src\ was first detected on December 18, 2005 by the SWIFT satellite during
an X-ray outburst \citep{palm05}. The source showed X-ray pulsations with
period $\sim$ 15 s and strong flaring activity. The spectrum was consistent
with a power law with $\Gamma\approx 3.2$, giving $F_X=6.1 \times 10^{-9}$
erg cm$^{-2}$ s$^{-1}$ in the energy range 15-100 keV. Subsequent
observations (see Table~\ref{xobs}) showed that \src\ is  a highly variable
source with a pulse period of 15.37682$\pm$0.00005 s  \citep{mark05} and
pulse fraction that increases from $\sim$50\% between flares to $\sim$80\%
during the flares \citep{bell06}. The X-ray spectrum is well fitted by an
absorbed power-law component and a high-energy cutoff. An iron line at 6.4
keV is also detected \citep{mark05,bell06}. However, the values of the
spectral parameters differ considerably between observations.
\citet{mark05} gave $E_{\rm cutoff}=5$ keV and $N_{\rm H}=2\times 10^{22}$
cm$^{-2}$, whereas \citet{bell06} gave  $E_{\rm cutoff}=12$ keV and $N_{\rm
H}=5\times 10^{22}$ cm$^{-2}$. \citet{camp06} did not detect the iron line
nor a cutoff. These authors reported a harder and less absorbed spectrum
($\Gamma=0.7$, $N_{\rm H}=0.95\times 10^{22}$ cm$^{-2}$). Finally, an
INTEGRAL spectrum in the range 4-60 kev obtained by \citet{tara06} gave
$\Gamma=0.6$, $E_{\rm cutoff}=6.7$ keV. The optical counterpart is believed
to be 2MASS16263652--5156305 \citep{rea06,nema06}, a Be star showing strong
H$_\alpha$ emission. A {\em Chandra} observation confirmed that the
position of the X-ray source is consistent with the suggested counterpart.
Chandra coordinates are RA: 16h26m36.5s, Dec: -51d56m30.7s, with an error
of 0.6 arcsec (Homan, private communication).

In this work we present the results of a timing and spectral analysis of
\src\ using RXTE data, limited to the three observations early in the outburst 
where flares are observed. 
The details of the observation are explained in
Sect.~\ref{obs} and a log presented in Table~\ref{tobs}. In Sect.~\ref{time}
the periodic and aperiodic time variability of \src\ is investigated.
Sect.~\ref{spec} deals with the time-average and pulse-phase spectral
variability. In Sect.~\ref{disc} a comparison of the timing and spectral
properties of  the flare and out-of-flare emission is performed and 
the nature of the system is discussed.

%------------------------------------------------------------------------------
\begin{table*}
\begin{center}
\caption{X-ray observations of \src\ in the literature and best-fit spectral 
parameters of an absorbed power law with cutoff.}
\label{xobs}
\begin{tabular}{ccccccccc}
\hline \hline \noalign{\smallskip}
Date	&Satellite 	&Flux$^a$&Energy range	&$\Gamma$ &$E_{\rm cut}$ &$N^b_{\rm H}$ &Fe line &Reference \\
	&		&	&(keV)		&	&	(keV)	&		&(keV)	&\\
\hline \noalign{\smallskip}
18-12-2005&SWIFT &6.1	&15-100	&3.2	&-	&-	&-	&\citet{palm05} \\
19-12-2005&RXTE	 &4.1	&2-40	&0.9	&5	&2	&6.4	&\citet{mark05} \\
19-12-2005&RXTE	 &3.6-2.4&3-25	&0.9-1.3&12	&4-5.5	&6.4	&\citet{bell06} \\
12-01-2006&SWIFT &1.2	&0.5-10	&0.67	&-	&0.94	&-	&\citet{camp06} \\
21-01-2006&INTEGRAL&1.6	&4-40	&0.6	&6.7	&-	&-	&\citet{tara06} \\
\hline \hline \noalign{\smallskip}
\multicolumn{9}{l}{$a$: $\times 10^{-9}$ erg cm$^{-2}$ s$^{-1}$}\\
\multicolumn{9}{l}{$b$: $\times 10^{22}$ cm$^{-2}$}\\
\end{tabular}
\end{center}
\end{table*}
%------------------------------------------------------------------------------

%------------------------------------------------------------------------------
\begin{figure}
\resizebox{\hsize}{!}{\includegraphics{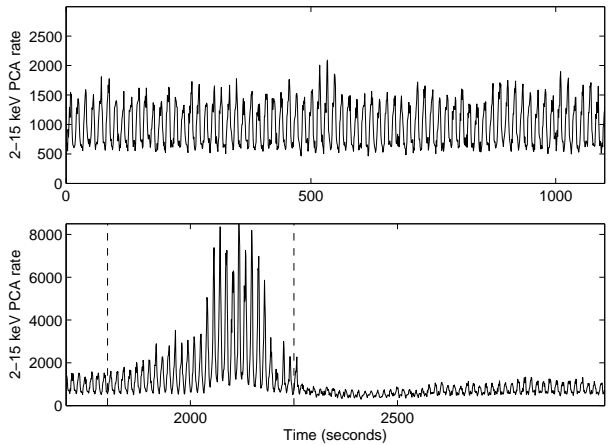} } 
\caption[]{Light curve of Obs. 1 
showing the X-ray pulsations and the increase of the pulse
fraction during the flare event. Bin size is 1 s. The dashed lines mark the 
flare interval used in the analysis (see text).}
\label{obs1}
\end{figure}
%------------------------------------------------------------------------------
%------------------------------------------------------------------------------
\begin{figure}
\resizebox{\hsize}{!}{\includegraphics{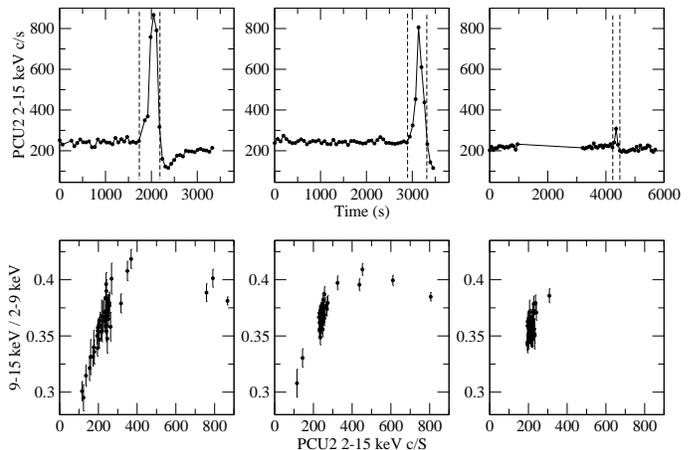} } 
\caption[]{Light curves and hardness-intensity diagrams of the 
three observations. The Y-axis was left the same in the three panels to allow 
easy comparison.  The entire time interval is used with a bin size of
64 s. The dashed lines indicate the limits of the
pre-, post- and flare intervals selected for the analysis.}
\label{lc_hr}
\end{figure}
%------------------------------------------------------------------------------

\section{Observations and data analysis}
\label{obs}

\src\ was observed with the Proportional Counter Array (PCA) onboard the
Rossi X-ray Timing Explorer (RXTE) starting from December 19,  2005, one
day after the discovery by Swift.  Here we concentrate on the three
observations made between December 24-27, 2005 (MJD 53728.385--53731.295),
in which the source displayed flaring activity. The data correspond to RXTE
proposal P91094. Four PCUs were on during the entire duration of the three
observation intervals. PCU1 was off all the time. 

Light curves and spectra were extracted using version v.5.3.1 of the RXTE
FTOOLS package. The following screening criteria were applied to the data
prior extraction: good time intervals were defined  when the pointing of
the satellite was stable ($<0.02''$), the elevation above 10$^{\circ}$ and
far away from the South Atlantic Anomaly.

The log of the observations is given in  Table~\ref{obs}.  
%We carried out
%an X-ray timing and spectral analysis of the three observation interval
%described in Table~\ref{obs}. 
Each observation contains one
flare event, although the event that took place in Obs. 3 was significantly
smaller and shorter. The duration of the flares varied from  $\sim$180 to 480
s. The flare in Obs. 2 was detected at the end of the observation
interval. Thus, the duration of post-flare interval of Obs. 2  is short
($\sim$ 100 s). The pre-flare interval of Obs. 3 contains a data gap of
about 2260 s due to the passage through the SAA anomaly. 

%There is a number of difficulties that hampered our analysis. The
%pre-flare data of Obs. 1 is affected by instrumental gaps due to data loss
%as a consequence of the failure of one of the transponders on September 1.
%1999.  

%------------------------------------------------------------------------------
\begin{table*}
\begin{center}
\caption{Summary of the observations.}
\label{tobs}
\begin{tabular}{cccc|ccc|ccc|ccc}
\hline \hline \noalign{\smallskip}
Obs.	 &RXTE ID &MJD    &MJD    &\multicolumn{3}{c}{Pre-flare}  &\multicolumn{3}{c}{Flare}&\multicolumn{3}{c}{Post-flare}\\
interval &	  &start  &end	&$I_X^a$ &Duration &PF &$I_{\rm peak}$$^a$ &Duration &PF &$I_X^a$ &Duration &PF  \\
\hline \noalign{\smallskip}
1 &91094-02-02-01 &53728.385 &53728.425      &995    &1800   &44\%   &3555   &450    &72\%   &750    &1142   &31\%\\
2 &91094-02-02-02 &53729.168 &53729.211      &988    &2900   &46\%   &3880   &480    &70\%   &600    &108    &36\%\\
3 &91094-02-02-04 &53731.228 &53731.295      &680    &2057   &38\%   &1040   &180    &51\%   &645    &1236   &37\%\\
\hline \hline \noalign{\smallskip}
\multicolumn{13}{l}{$a$: Background-subtracted count rate for 4 PCU in the energy
range 2-15 keV for a time bin  of 16 s.}\\
\multicolumn{13}{l}{Durations are in seconds.}\\

\end{tabular}
\end{center}
\end{table*}
%------------------------------------------------------------------------------
%------------------------------------------------------------------------------
%\begin{table*}
%\begin{center}
%\caption{Definition of pre-flare, flare and post-flare intervals in MJD.}
%\label{tint}
%\begin{tabular}{cccc}
%\hline \hline \noalign{\smallskip}
%Obs.	 &Pre-flare  			&Flare			&Post-flare\\
%\hline \noalign{\smallskip}
%1 	&53728.384252--53728.405085	&53728.405085--53728.410295	&53728.410295--53728.423513 \\ 
%2 	&53729.169264--53729.202828	&53729.202828--53729.208384	&53729.208384--53729.209634 \\
%3 	&53731.228279--53731.239531	&53731.278049--53731.280133	&53731.280133--53731.294438 \\
%	&53731.265490--53731.278049	&				&	\\
%\hline \hline \noalign{\smallskip}
%\end{tabular}
%\end{center}
%\end{table*}
%------------------------------------------------------------------------------

\section{Timing analysis}
\label{time}

\subsection{Light curves and hardness ratios}

Strong pulsations are detected at the 15.37s period from all observations.
Figure~\ref{obs1} shows the light curve of Obs. 1 at high resolution
($\Delta t=1$ s). The top panel shows the first 1.2 ks of data, where the
pulsation is evident. The bottom panel shows the second part of the
observation, which includes a strong flare. Its shape is very similar to
that of the other strong flare (from Obs. 2): a major increase in the
pulsed fraction, clearly accompanied by a more moderate increase of the DC
level, followed by a depression of DC level and pulsed fraction and a slow
recovery. Fig.~\ref{lc_hr} shows the three flare
events with corresponding hardness-intensity diagrams using a coarser time
bin ($\Delta t=64$ s). The hardness ratio was defined as the ratio between
the count rate in the 9-15 keV band to the count rate in the 2-9 keV band.
The flare in Obs. 2 is very similar to that in Obs. 1, although it appears
later in the observation and the post-flare region is not observed. The
flare in Obs. 3 is much weaker.  As the count rate increases, the
hardness ratio increases until the count rate reaches about 350 c s$^{-1}$
at which point it flattens. The dashed lines in Fig.~\ref{obs1} and
\ref{lc_hr} indicate our definition of pre-flare, flare and post-flare
intervals for each observation. 

%Given the weakness of the third flare and the incompleteness of the second,
%we concentrate most of our flare analysis  on the first flare, from Obs. 1.

As mentioned, during the flares, the pulsed fraction increases
considerably; the minima of the pulsation cycle increase by a factor of 2,
while the maxima increase by a factor of 5. After the flares, both the
average rate and the pulsed fraction go below their pre-flare values, and
start a slow recovery. In Obs. 1 the flux had not completely recovered the
pre-flare level 1200 seconds later, when the observation ended.   Given
the fact that the pre-flare level of Obs. 2 is similar to that of Obs. 1,
it is likely that the source intensity fully recovered to the pre-flare
level.

%This means that at the peak of the flare the emission is softer than during the
%rise or decay. 

In order to investigate  differences in the emission properties between the
flare and quiescent states, we isolated the time intervals corresponding
to the flare events. For each observation, we defined three intervals:
pre-flare, flare and post-flare. The exact definition of these intervals
can be seen in Fig.~\ref{lc_hr}.

The average count rate (peak intensity in the case of the flare interval),
duration and pulse fraction of the available intervals corresponding to the
three "states" (pre-flare, flare and post-flare) for each observation are
given in Table~\ref{tobs}.

\subsection{Spin period determination}
%\subsubsection{Non-flare emission}

%The spin period determination was performed with a 0.125-s binned light
%curve, extracted from the {\it B\_2ms\_8B\_0\_35\_Q} mode, which provides 2
%ms timing resolution and covers the energy range 2-15 keV.

In order to determine the spin period of \src\ we obtained a 0.125-s binned
barycentric corrected light curve of the non-flaring emission for the three
observation intervals in the energy range 2-15 keV (from PCA {\it
B\_2ms\_8B\_0\_35\_Q} mode). An FFT applied to the light curve revealed a
coherent modulation at 0.065 Hz, which correspond to a pulse period of 15.4
s. Several of its harmonics are clearly seen. In order to accurately
measure the pulse period, a pulse-folding analysis was performed. The light
curve was divided into blocks of continuous stretches of data points with
a duration of 230 s. Each block was folded modulo the initial period found
from the FFT. The resulting pulse profile obtained for the first block was
used as a template. The difference between the actual pulse period at the
epoch of observation and the period used to fold the data was determined by
cross-correlating the pulse profiles corresponding to each block and the
template. The phase shifts were fitted to a linear function, whose slope
represents the correction in frequency  to the actual period. A new
template was derived for this corrected period, and the process repeated.
The derived pulse periods were 15.3718$\pm$0.0003 s, 15.3711$\pm$±0.0002 s
and 15.3724$\pm$0.0007 s for the three observations, respectively.

\subsection{Pulse profile and pulse fraction}

\begin{figure}
\resizebox{\hsize}{!}{\includegraphics{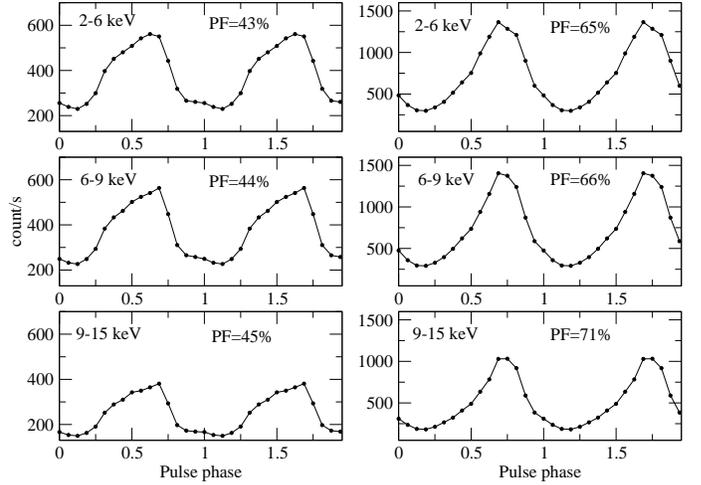}} \\
\caption[]{Comparison of the pulse profile as a function of energy during the
flare (right) and non-flaring (left) part of the light curve. Data correspond 
to Obs 1.}
\label{pp}
\end{figure}

The average pulse profiles at three different energy ranges, obtained by
folding the corresponding light curves onto the best-fit spin period, are
shown in Fig.~\ref{pp}. The pulse profile of the flare is somehow narrower
than that of the out-of-flare but for both there is no strong dependence on energy.
The pulse fraction was calculated as PF
=($I_{max}$--$I_{min}$)/($I_{max}$+$I_{min}$).

%Figure~\ref{pf_time} shows the pulse fraction (defined as PF
%=($I_{max}$--$I_{min}$)/($I_{max}$+$I_{min}$)) corresponding to each cycle
%as a function of time for Obs. 1. Note the increase of the PF during the
%flare event.

%of the non-flare emission was obtained by folding the
%light curve the best-fit spin period. The resulting pulse profiles as a
%function of energy for Obs. 2 are shown in Fig.~\ref{pp}.  The pulse
%profile of the flare is narrower than that of the out-of-flare and the pulsed
%fraction (defined as PF =($I_{max}$--$I_{min}$)/($I_{max}$+$I_{min}$))
%increases only moderately with energy.
%For the flare in Obs. 1, the pulse fraction increases from $\sim$45\%
%before the flare to $\sim$70\% during the flares, as is clearly seen in
%Fig.~\ref{pf_time}, where the pulse fraction for each cycle as a function
%of time for Obs. 1 is shown.  After the flare, not only the count rate but
%also the pulse fraction display lower values than before the flare
%($\sim$35\%).

%Figure \ref{profvar} shows the variability of the pulse profile as a
%function of the pulse phase. The largest variability occurs near the peak
%of the pulse profile. The rms was calculated as $rms=\sigma/flux$, where is
%the standard deviation of the absolute value of the difference between
%every point of the light curve and the corresponding average point of the
%pulse profile and flux is the average count rate of the corresponding phase
%bin. CHECK THIS AGAIN. The light curve was binned so that one pulse
%contains 32 phase bins.

%------------------------------------------------------------------------------
%------------------------------------------------------------------------------

\begin{figure}
\resizebox{\hsize}{!}{\includegraphics{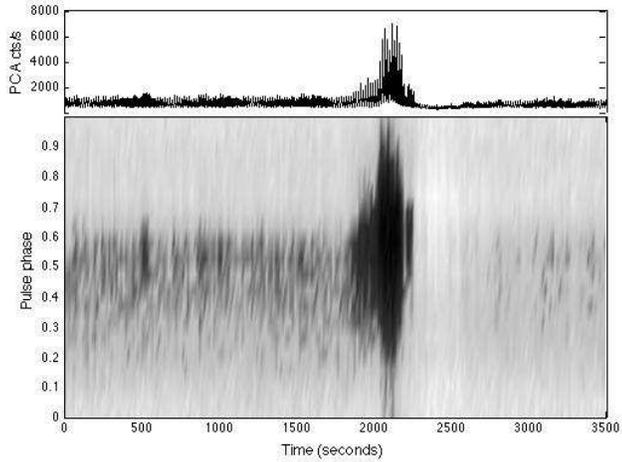}} \\
\caption[]{Top panel: light curve of Obs. 1. Bottom panel: corresponding 
time-phase image, obtained aligning all pulses as a function of time. 
Darker areas indicate higher count rate.
}
\label{pulse_image}
\end{figure}

In order to perform a more in-depth analysis of the individual pulse
profiles, we extracted a light curve from the full PCA energy range with  a
bin size of $P_{\rm spin}/20$, divided it in intervals 20-points long
(centered approximately on the pulse maximum) and with these constructed a
time-phase image, with time on the X-axis, pulse phase on the Y-axis and
count rate on the Z axis. The image on  a gray scale color map is shown in
Fig. \ref{pulse_image}, together with the corresponding light curve (top
panel). From this figure, one can see that the out-of-flare pulse peaks at
around phase 0.5, while the flare itself is characterized by a maximum
around phase 0.6. Moreover, at T$\sim$500 one can see a small rate increase
corresponding to a slightly higher phase, indicating that a very small
flare is probably present there.

In order to study in more detail the pulse changes during the flare, we
accumulated an out-of-flare template by simply adding the first 100 cycles.
Comparing this template with the single pulses during the flare, it appears
as if the flare could be characterized by the appearance of one or two
additional components at fixed phases.

%------------------------------------------------------------------------------
%------------------------------------------------------------------------------

In order  to assess this numerically, we ran the following procedure.  Each
cycle was fitted with the out-of-flare template with a variable overall 
multiplicative factor. Not surprisingly, outside  the  flare  and
post-flare  region  (and  with  the exception  of  three  cycles  around
cycle 35),  this  model  gives a decent  fit  with  normalization  equal 
to  1. By {\it decent} fit we mean a fit with a $\chi_r^2=\chi^2$/20 $<$15:
this is not formally acceptable, but no systematic residuals  are seen and
given the high statistics it can not be expected that a single simple model
could fit all single cycles. Overall, out of flare the $\chi^2_r$ values
are around 8.

For  the cycles with $\chi^2_r>15$,  we identify  the  largest  positive
deviation  from the template fitting (limited  to  the  phase  range 
0.2-0.8)  and  repeat  the  fit  with  the  addition  of  a  Gaussian 
component  centered  at  the  position  of  that  maximum  deviation, 
with  normalization  equal to the maximum deviation and with starting width
0.1. With this addition, only a few cycles around the maximum of the flare
retain a $\chi^2_r>15$.

\begin{figure}
\resizebox{\hsize}{!}{\includegraphics{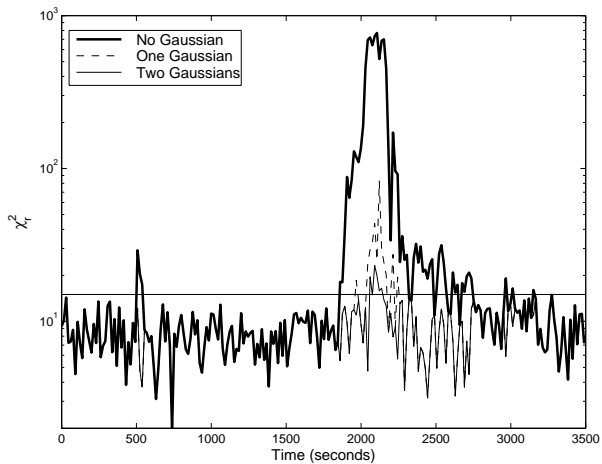}} \\
\caption[]{
Values of $\chi^2_r = \chi^2/20$ obtained with the fitting procedure described 
in the text. The different curves correspond to simple template (dotted line), 
template plus one Gaussian (dashed line) and template plus two Gaussians 
(full line). The horizontal line marks the level $\chi^2_r = 15$.
}
\label{chi2}
\end{figure}

In  order  to try  to find a  better  fit for  these cycles, we added an 
additional  Gaussian  component. The  procedure is the  same as  before. 
The new Gaussian  is  added  only  if  the  fit  with  a  single  Gaussian 
results  in  a  $\chi^2_r > 15$.  The fits are considerably  better  during
the flare, although there are still a  few points with a $\chi^2_r > 15$.
The $\chi^2_r$ values obtained with this procedure are shown in Fig.
\ref{chi2}.

%------------------------------------------------------------------------------
%------------------------------------------------------------------------------

\begin{figure}
\resizebox{\hsize}{!}{\includegraphics{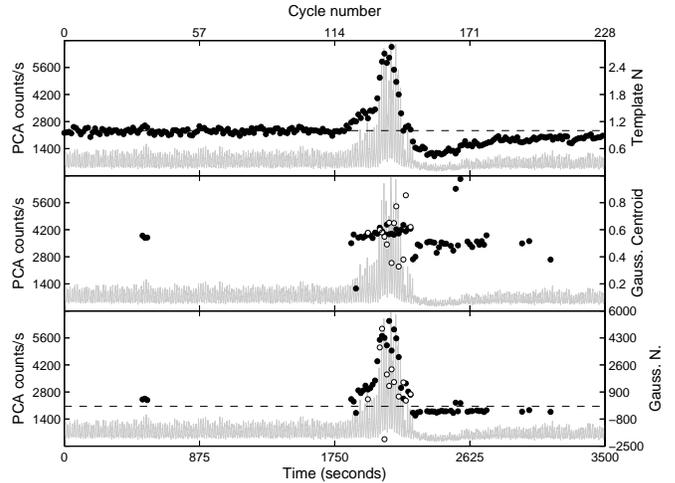}} \\
\caption[]{
Best fit parameters as a function of time. On all three panels, the gray curve shows the PCA count rate evolution for reference. Top panel: multiplicative factor for the out-of-flare template. The dashed line shows the level of 1. Middle panel: phase of the centroid of the Gaussians (first Gaussian: filled circles; second Gaussian: empty circles).
Bottom panel: normalization of the Gaussians (first Gaussian: filled circles; second Gaussian: empty circles). The dashed line corresponds to zero normalization.
}
\label{parameters}
\end{figure}

The final fits have a number of free parameters between one (the template
normalization factor for the template-only fits) and seven (template
normalization plus three parameters for each of the two Gaussians) for the
most complex fits. The best fit values for five of the seven parameters are
shown in Fig. \ref{parameters}. The remaining two parameters, the width of
the two Gaussians, are clustered around 0.1-0.2 and 0.01-0.05 in phase for
the first and second Gaussian respectively.

%------------------------------------------------------------------------------
%------------------------------------------------------------------------------

Based on these results, we can identify three intervals:
\begin{itemize}

\item Before the flare: the pulse shape is consistent with a constant
shape, corresponding to what we used as template.

\item During the flare: the template shows a higher normalization factor
(up to 3), while an additional broad Gaussian centered at phase 0.6
appears. Occasionally, a second narrower Gaussian is necessary for a fit.

\item During the post-flare depression: the template normalization is lower
than unity and a negative Gaussian with a very constant normalization is
apparent.

\item After the post-flare depression: the template normalization slowly
recovers to unity, no Gaussians are needed.

\end{itemize}

%------------------------------------------------------------------------------
\begin{figure}
\resizebox{\hsize}{!}{\includegraphics{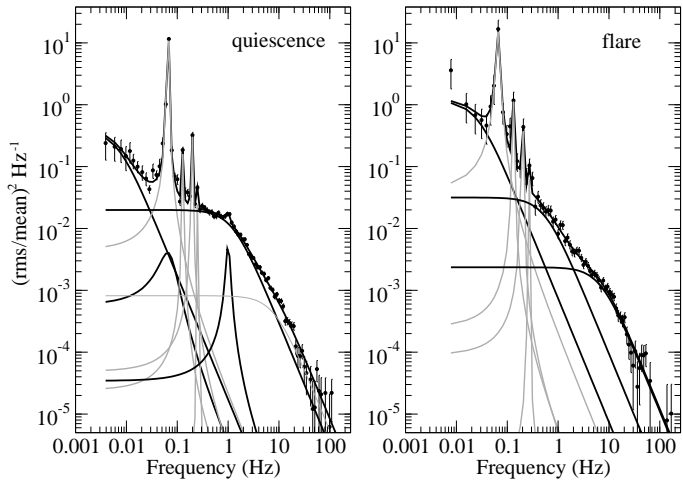}} \\
\caption[]{Power spectrum of \src. Coherent pulsations are clearly seen at
$\sim$0.065 Hz together with some higher harmonics. The quiescence emission 
shows a weak QPO components at 1 Hz. The lines indicate the Lorentzian 
components used to fit the power spectra (see text)}
\label{psd}
\end{figure}
%------------------------------------------------------------------------------

\subsection{Power spectrum and noise components}

In order to obtain the power spectra, the Obs. 2 light curve was binned
with bin size $2^{-9}$ s, which is the maximum resolution provided by the
{\it B\_2ms\_8B\_0\_35\_Q} data mode. Then the light curve was divided into
segments of 512 s for the non-flare part and 128 s for the flare. An FFT
was calculated for each segment (6 for the non-flare and 4 for the flare).
The final power spectra resulted after averaging all the individual power
spectra and rebinning in frequency. Figure~\ref{psd} shows the power
spectra and noise components of the preflare and flare light curves.  In
addition to the harmonic components of the pulsations, strong band-limited
noise is detected both in the quiescence (i.e. non-flaring emission) and
during the flare, with an integrated 0.01-100 Hz fractional rms of around
40\% and 64\%, respectively.  The change in the pulse profile between 
pre-flare and flare (as seen in Fig.~\ref{pp}) can also be expected from the
change in the relative strength of the harmonics of the pulsation. 

The power spectra were fitted with Lorentzians. Ignoring the X-ray
pulsation and its harmonics, a total of 5 Lorentzians were required to fit
the non-flare power spectrum, while the flare power spectrum needed 3
Lorentzians.

Three Lorentzians are common to the two data sets. These are zero-centred
Lorentziants that fit the continuum. The power spectra of the non-flare
part require two more Lorentzians. One is centred at the same frequency as
the pulsation and can be interpreted as the coupling between the periodic
and aperiodic variability \citep{lazz97}. This coupling can be recognised by
the broadening of the wings of the narrow peaks of the periodic modulation.
The lack of this component in the flare power spectrum maybe due to the
fact that the aperiodic noise is stronger (see the higher continuum
short-ward of the pulsation) during the flare.
The non-flare power spectrum presents a QPO at around $\nu=1$ Hz with Q
($\nu$/FWHM) of 4 and fractional rms amplitude of 4.7\%.

%------------------------------------------------------------------------------
\begin{figure}
\includegraphics[width=8cm,height=5cm]{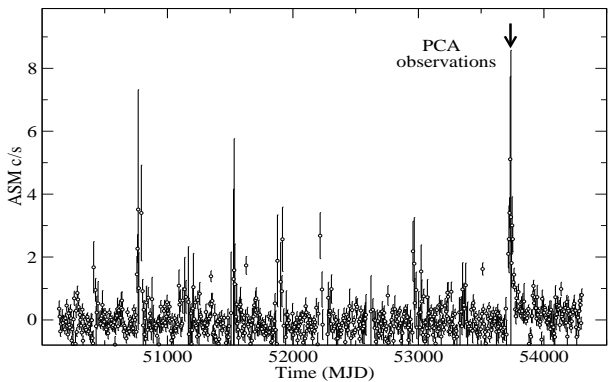} 
\caption[]{5-d binned ASM light curve of \src\ for the period 
MJD 50135--54340. The start time of the PCA observations is 
indicated by an arrow. }
\label{asm}
\end{figure}
%------------------------------------------------------------------------------
%------------------------------------------------------------------------------
\begin{figure}
\includegraphics[width=8cm,height=5cm]{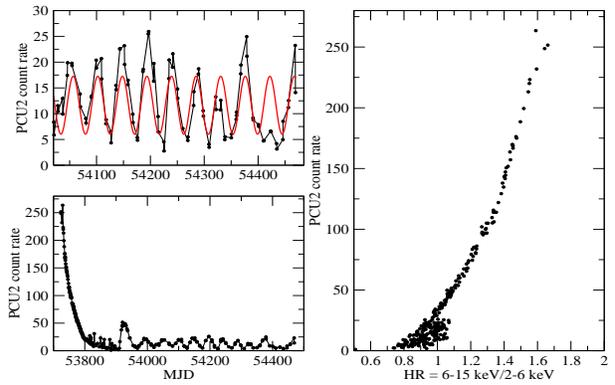} 
\caption[]{After the outburst that led to
its discovery as an X-ray source, \src\ exhibited a periodic
modulation with period $\sim$45 days. On the right the hardness
ratio-intensity diagram is shown.}
\label{outburst}
\end{figure}
%------------------------------------------------------------------------------

\subsection{Longer-term variability}

 Figure~\ref{asm} shows the long-term variability of \src\ and the
transient nature of its X-ray emission. The source spends most of the time
in a quiescent state (the average count rate is $\approx$0.06 ASM c
s$^{-1}$) and only occasionally goes into outburst. Note that significances
above 3 sigma are only seen in the 5-day average ASM bins that cover the
2005 December outburst. 

We also collected the average PCU2 count rates for each of the 295 RXTE
observations, thus resulting in an unevenly spaced time series (see
Fig.~\ref{outburst}). The onset of the PCA observations coincided with the
outburst peak (Fig.~\ref{asm}). A first visual inspection of the resulting
lightcurve shows two main properties: {\em i)} an initial fast flux decay
lasting for about 160 days since the outburst onset, followed by {\em ii)}
a series of apparently periodic "bumps".  In order to carry out a first
estimate of the two components we fitted the first 160 day points with an
exponential component and the second part of the lightcurve with a
sinusoid. For the first component we found that $\tau$ of the exponential
is $\sim$33 days. For the second component we obtained a period of  45$\pm$1
days, though the reduced $\chi^2$ is only about 2. 

%This period might be ascribed to the orbital motion of the accreting
%neutron star around its companion.

We note that bumps are rather asymmetric, suggesting either that the
sinusoid model is a rough estimate of the raw data and/or the 45-d
period is a harmonic of the true period. This interpretation is also
corroborated by the very high asymmetric profile of the latest four peaks,
with the odd and even bumps differing each other. To test this hypothesis
we add an additional sinusoidal component to the fit with a period forced
to be twice to that of the first component. The inclusion of this component
has a probability of $\sim$99.0\% to be significant. It is evident that
only future optical spectroscopic studies of the companion star and/or
pulse arrival time analysis of the X-ray photons can provide an unambiguous
detrmination of the orbital period.  The first bump after
the exponential flux decay phase is a reminiscence of what is routinely
observed in many Be/X-ray binaries as they enter in an outburst phase
\citep[see e.g.][]{camp99}: a rapid rise in flux, a flat topped maximum and
a relatively slow decay. At variance with the above interpretation is the
second bump which falls at the minimum of the expected modulation. 

Figure~\ref{outburst} also shows the variation of the hardness of the
source, defined as the ratio between the count rates in the energy bands
6-15 keV to  2-6 keV, as a function of intensity. As the outburst decayed
the source became softer .

It is worth mentioning that the same flare mechanism that occurs during
high-intensity states, i.e., near the outburst peak, still works at lower
mass accretion rates. The variability seen at the end of the outburst
decay, around MJD 53800-53900 (see Fig.~\ref{outburst}), is caused by the
same type of flaring activity that we have reported in previous sections.
The flares show the same factor 3-4 increase in flux and also an increase
in the pulse fraction.

%----------------------------------------------------------------------------
\begin{table}
\begin{center}
\label{psdfit}
\caption{Results of the fits to the power spectra. Only the non-coherent
components are shown.}
\begin{tabular}{lcc}
\hline
Lorentzian		&quiescence		&flare	\\
\hline \hline
$\nu_1$	(Hz)		&0$^f$			&0$^f$	\\
$FWHM_1$ (Hz)		&0.013$\pm$0.007	&0.012$\pm$0.009	\\
rms$_1$	(\%)		&7.8$\pm$1.3		&38.7$\pm$9.0	\\
$\nu_2$			&0.065$^f$		&--	\\
$FWHM_2$		&0.4$^{+3.4}_{-0.2}$	&--	\\
rms$_2$			&5.4$\pm$1.9		&--	\\
$\nu_3$			&1.02$\pm$0.03		&--	\\
$FWHM_3$		&0.23$\pm$0.07		&--	\\
rms$_3$			&4.7$\pm$1.0		&--	\\
$\nu_4$			&0$^f$			&0$^f$	\\
$FWHM_4$		&2.4$\pm$0.2		&1.0$\pm$0.3	\\
rms$_4$			&26.2$\pm$1.0		&24.9$\pm$1.6	\\
$\nu_5$			&0$^f$			&0$^f$	\\
$FWHM_5$		&13.7$\pm$1.3		&13.5$\pm$1.7	\\
rms$_5$			&15.9$\pm$1.0		&22.3$\pm$0.9	\\
\hline \hline 
\multicolumn{3}{l}{$f$: fixed} \\
\end{tabular}
\end{center}
\end{table}
%------------------------------------------------------------------------------

\section{Spectral analysis}
\label{spec}

\subsection{Time-averaged spectroscopy}

\begin{figure}
\includegraphics[width=8cm,height=5cm]{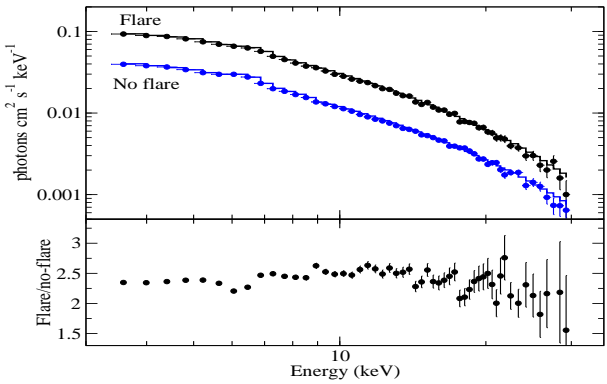}
\caption[]{Energy spectra of the flare and out-of-the-flare parts for Obs
1. The lower panel shows the ratio between the two spectra.}
\label{sp}
\end{figure}

In order to investigate potential spectral variability between the
non-flare emission and the emission during the flare, we extracted energy
spectra corresponding to the flare and a stretch of non-flare emission.
Fig.~\ref{sp} shows the spectra for Obs.1 and Table~\ref{model} summarises
the results of the spectral fits for the three observations. A systematic
error of 1\% was added in quadrature to the statistical one. An absorbed
power law with a high-energy cut off and an iron line at $\sim6.4$ keV
provided a good fit to the spectral continua. No spectral variability is
seen within the uncertainties, although the flare spectrum in Obs. 2 is
somehow harder than the preflare one. The pre-flare and flare spectra
basically  differ in that the  pre-flare spectra require the inclusion of
an edge at $\sim$9 keV to obtain an acceptable fit. The probability that
the improvement of the fit occurs by chance (F-test) by the addition of
this component is $\simless 10^{-5}$. 

%----------------------------------------------------------------------------
\begin{table*}
\begin{center}
\label{model}
\caption{Spectral fits results for the three observations. Uncertainties are 
90\% confidence. Flux corresponds to the energy range 3-30 keV. }
\begin{tabular}{lcc|cc|cc}
\hline
				&\multicolumn{2}{c}{Obs. 1}					&\multicolumn{2}{c}{Obs. 2}		&\multicolumn{2}{c}{Obs. 3} \\
\hline
parameters			&pre-flare			&flare			&pre-flare		&flare			&pre-flare		&flare\\
\hline \hline
Flux (erg cm$^{-2}$ s$^{-1}$)	&$3.5 \times 10^{-9}$		&$8.4 \times 10^{-9}$	&$3.5 \times 10^{-9}$	&$6.3 \times 10^{-9}$	&$3.1 \times 10^{-9}$	&$3.8 \times 10^{-9}$	\\
$N_H$ ($\times 10^{21}$ cm$^{-2}$)&9.4$^*$			&9.4$^*$		&9.4$^*$		&9.4$^*$	  	&9.4$^*$		&9.4$^*$	\\
$\Gamma$			&0.73$^{+0.08}_{-0.05}$		&0.6$\pm$0.1		&0.8$\pm$0.1		&0.5$\pm$0.1	 	&0.82$\pm$0.06		&0.8$\pm$0.1	\\
norm.				&0.11$\pm$0.01			&0.22$\pm$0.03		&0.12$\pm$0.01		&0.15$\pm$0.02		&0.12$\pm$0.01		&0.13$\pm$0.02	\\
$E_{cut}$	(keV)		&4.7$\pm$0.3			&4.8$\pm$0.4		&5.0$^{+0.1}_{-0.4}$	&4.5$\pm$0.4 		&5.0$\pm$0.3		&5.0$\pm$0.4	\\
E$_{fold}$ (keV)		&9.7$^{+0.8}_{-0.4}$		&8.6$\pm$0.6		&10.3$\pm$0.5		&8.2$^{+0.6}_{-0.3}$ 	&10.0$^{+0.8}_{-0.5}$	&10.6$^{+1.4}_{-0.8}$	 \\
$E_{\rm Fe}$ (keV)		&6.2$\pm$0.1			&6.5$\pm$0.2		&6.3$\pm$0.1		&6.4$\pm$0.1		&6.3$\pm$0.1		&6.4$\pm$0.1	 \\
$\sigma_{\rm Fe}$ (keV)		&0.4$^*$			&0.4$^*$		&0.4$^*$		&0.4$^*$		&0.4$^*$		 &0.4$^*$	\\
norm. ($\times 10^{-3}$)	&6$\pm$1			&8$\pm$3		&5.2$^{+1}_{-0.6}$	&9$\pm$2		&5$\pm$1		&7$\pm$1	\\
EW(Fe) (eV)			&195$\pm$30			&105$\pm$30		&180$\pm$30		&160$\pm$20		&195$\pm$35		&235$\pm$80	\\
$E_{\rm edge}$ (keV)		&8.7$^{+0.8}_{-0.4}$		&--			&8.9$\pm$0.5		&--	    		&8.9$\pm$0.5		&-- 	\\
$\tau_{\rm edge}$		&0.07$\pm$0.03			&--			&0.07$\pm$0.02		&-	     		&0.06$\pm$0.02		&--	\\
$\chi^2_r$(dof) 		&0.98(46)			&0.85(48)		&0.99(46)		&0.91(48)   		&0.82(48)		&1.15(48)	\\
\hline \hline 
\multicolumn{3}{l}{*: fixed} \\
\end{tabular}
\end{center}
\end{table*}
%------------------------------------------------------------------------------

%------------------------------------------------------------------------------
%------------------------------------------------------------------------------

\subsection{Pulse-phase spectroscopy}

As mentioned above, the shape of the pulse profile of \src\ hardly
changes with energy. A more detailed way to investigate how the X-ray
spectrum changes as a function of the spin of the magnetized neutron star
is by performing pulse-phase spectroscopy. In order to investigate the pulse
phase dependence, high time and energy resolution are needed
simultaneously. Unfortunately, the only data mode available in our
observations that provides detailed time and energy information is {\it
B\_2ms\_8B\_0\_35\_Q}, which provide high time resolution but modest energy
resolution, namely, eight energy channels in the energy range 2.6-15 keV.

\begin{figure}
\includegraphics[width=8cm,height=5cm]{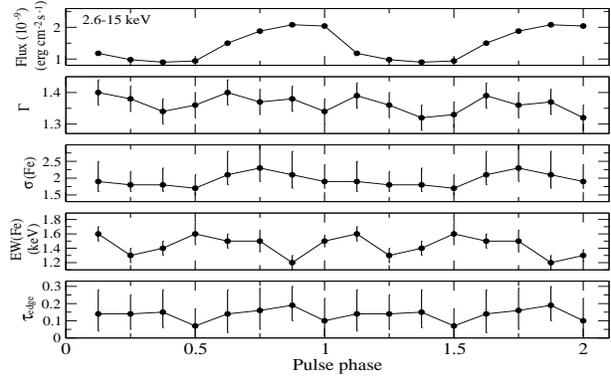} 
\caption[]{Spectral parameters as a function of pulse phase for the
preflare state of Obs. 2.}
\label{phasefit}
\end{figure}

The pulse profile of Obs. 2 was divided into eight equally spaced bins to
provide eight 2.6-15 keV background-subtracted spectra. Each spectrum was
fitted with a power law plus iron line and edge. Due to the modest energy
resolution, the iron line parameters were not well constrained. The iron
line and edge energy were fixed to 6.4 keV and 8.8 keV, respectively, the
value obtained from the time-average spectrum. As can be seen in
Fig.~\ref{phasefit}, despite that the flux changed by a factor 2, no
significant changes were seen in any of the spectral parameters, neither in
the continuum (power-law photon index) nor in the line and edge (width,
$\tau_{\rm edge}$). This result confirms the absence of variability of the
pulse profiles with energy.

%------------------------------------------------------------------------------
%------------------------------------------------------------------------------

\section{Discussion}
\label{disc}

We have performed a timing and spectral analysis of the newly discovered
X-ray pulsar \src. We have concentrated on the observations made at the
early phase of the reported outburst (see Fig.~\ref{asm} and
\ref{outburst}) in which the source displays flaring activity. Each X-ray
light curve was divided into subintervals corresponding to the out-of-flare
and flare parts of the observations. A separate analysis was carried out in
each subinterval.

\subsection{Out-of-flare analysis}

The out-of-flare parts of the light curves were used to derive the spin
period because these parts do not present high amplitude flux changes and 
because of their longer duration, allowing enough number of cycles to be
included. No significant variations in the spin period were seen during the
observations. Thus we derived a weighted mean spin period of
15.3714$\pm$0.0003 s. 

The pulse profile of the out-of-flare part of the X-ray light curve,
obtained by considering 100 cycles folded on the spin period, and a
variable overall  multiplicative factor fits well the individual pulses
with  normalization  equal  to unity. In the case of the post-flare
interval, where  both  the  overall  flux  level  and  the  pulse 
amplitude  are  depressed, the best fits  are  obtained with  the 
addition  of a negative Gaussian  component.

The average non-flare X-ray flux in the 3-30 keV band is $3.5\times
10^{-9}$ erg s$^{-1}$ cm$^{-2}$ for Obs. 1 and Obs. 2 and slightly lower in
Obs. 3, $2.9\times 10^{-9}$ erg s$^{-1}$ cm$^{-2}$. 
 There is no clear evidence for variability of the spectral parameters
with time nor between the flare and out-of-flare spectra in any of the
three observations. The only difference is the presence of an absorption
edge component in the out-of-flare spectrum that it is not required in the
flare spectrum.

\subsection{Flare analysis}

The flare event consists of a sudden increase of the X-ray flux by a factor
of four and is accompanied by an increase of the pulse fraction and a
hardening of the spectrum. The average 3-30 keV flux, taking into account
the entire duration of each flare, decreased from  $8.5\times 10^{-9}$ erg
s$^{-1}$ cm$^{-2}$ in Obs. 1 to $6.3\times 10^{-9}$ erg s$^{-1}$ cm$^{-2}$
in Obs. 2 and further to $3.8 \times 10^{-9}$ erg s$^{-1}$ cm$^{-2}$ for
Obs. 3.

The template fitting showed that the flare corresponds to the appearance of
an additional component at a  different phase than  the ``normal" pulse shape
outside flares, in addition to a brightening of the out-of-flare template. 
The average pulse profile obtained before the flare does not generally fits
well the individual pulses during the flare. The addition of a Gaussian 
component centered around phase 0.6 (shifted from the peak at phase 0.5 
of the template shape) reduces largely (see Fig. \ref{chi2}) the amount of bad fits.
Still, there are some pulses that do not fit well to an average profile
plus a Gaussian. The addition of a second Gaussian improved further 
the fits during the flare. In the post-flare phase, when the template normalization
goes below unity, this additional component disappears and is replaced by a 
{\it negative} Gaussian component centered at phase 0.5. As the template
slowly recovers to pre-flare levels, the negative component also disappears.

This behavior is difficult to understand. The flare appears to be made of
a general increase in the pulse fraction {\it plus} the appearance of a peaked
component at phase 0.6. The relatively small difference in average energy 
spectrum between non-flare and flare indicates that the spectrum of the
additional component is not very different from that of the template emission.
The post-flare negative contribution shows that after the flare the additional
Gaussian disappears and is replaced by a depression of the template maximum, 
together with a decrease of the pulse fraction. 

\subsection{The nature of the source}

The X-ray spectral components of \src, namely, absorbed power-law component
plus high-energy cutoff and iron line at 6.4 keV are characteristics of
accreting X-ray pulsars \citep{whit83,reig99,cobu02}.  Although a few
low-mass X-ray pulsars are known (e.g. Her X--1) the vast majority of
pulsating neutron star are found in high-mass X-ray binaries.  Massive
X-ray binaries are classified according to the luminosity class of the
optical component into supergiant X-ray binaries and Be/X-ray binaries. The
former tend to be persistent sources, have short ($P_{\rm orb} < 10$ d)
orbital periods and long spin periods ($P_{\rm spin} > 100$ s), while
Be/X-ray binaries are usually transients, have wider orbits ($P_{\rm orb} >
20$ d) and contain fast rotating neutron stars  ($P_{\rm spin} < 100$ s).  
On long timescales Be/X-ray binaries display two types of X-ray activity:
regular, orbitally modulated outbursts normally peaking at or close to
periastron  with X-ray flux increases of about one order of magnitude with
respect to the pre-outburst state, reaching peak luminosities $L_x \leq
10^{37}$ erg s$^{-1}$ (type I) and bright (X-ray flux  $10^{3}-10^{4}$
times that at quiescence) uncorrelated with orbital phase and that last for
several orbital periods (type II).

The long-term variability of \src\ shown in Fig.~\ref{asm} and
\ref{outburst} is reminiscent of Be/X-ray binaries \citep{reig07},
especially, the detection of smaller periodic outbursts (type I) following
a major one (type II). This behaviour has been seen in the Be/X-ray
binaries 4U\,0115+63 \citep{negu98},  KS 1947+300 \citep{gallo04} and EXO
2030+375 \citep{reig08}. The periodicity is associated with the orbital
period of the system. The X-ray emission increases as a result of accretion
from the circumstellar disk of the Be star during periastron passage.
Therefore, it seems natural to assume that the small outbursts in \src\ are
also separated by the orbital period. Nevertheless, given the asymmetric
profiles of the outbursts (note the last four shown in
Fig.~\ref{outburst}), we cannot exclude that the true orbital period is
twice the modulation. In fact, the optical counterpart to \src\
exhibits strong H$\alpha$ emission. \citet{nema06}  reported an H$\alpha$
equivalent width of $\sim$ --40 \AA.  Such equivalent width  implies an
orbital period $100 < P_{\rm orb} < 200$ days \citep{reig07}. Although the
H$\alpha$ line may also appear in emission in supergiant binaries, the
equivalent width is normally below $\sim$7 \AA\ in these systems. 

We can then conclude that all the observational data available of \src\
both from the vicinity of the compact object (relatively short spin
period, X-ray spectral parameters, transient X-ray emission) and the
optical companion  (strong H$\alpha$ emission, relatively long orbital
period) suggest the presence of a neutron star orbiting a Be star
companion. 

If the nature of \src\ as a Be/X-ray binary is finally confirmed, then this
is the first time that such superfast  flaring activity is reported for
this type of systems. Up to now the fastest X-ray variability seen in
Be/X-ray binaries was that associated with the rotation of the neutron star
(X-ray pulsations). 
Flaring activity is not uncommon in massive X-ray binaries habouring
supergiant companions (SGXR) but absent, with only one exception (EXO
2030+375), in Be/X-ray binaries.  During an X-ray outburst of the Be/X-ray
binary EXO 2030+375 detected by EXOSAT in 1985 October, a series of six
flares that recurred quasi-periodically every 3.96 hr were observed. The
duration of these flares varied between 1.3-2.3 hr and showed significant
variability during the decay, consisting of intensity quasi-priodic
oscillations with periods of 900-1220 s \citep{parm89}.

The flares of EXO 2030+375 share similar properties with those seen in LMC
X-4 flares, including the $\sim$3-4 hr flare recurrence time, the $\sim$1-2
hr flare duration, the high fluences of the flares ($>10^{40}$ ergs), the
rapid rise with gradual decay, and the absence of any significant spectral
change \citep{moon03}. These similarities suggest a common origin. One
possible mechanism could be Rayleigh-Taylor instabilities near the
magnetospheric boundary of the neutron star \citep{appa91}.

Flaring activity has also been reported in two wind-fed supergiant X-ray
binary. 4U 1907+09 exhibits flares on time scales of 1--2 hr
\citep{frit06}. These flares are locked to the orbital motion as they occur
twice per orbit \citep{zand98,muke01}. Vela X--1 has been observed to flare
above its persistent level on several occasions \citep{laur95,kriv03}. The
duration of the flares was 5-6 hr.

However, the duration of the flares in the systems mentioned above is
longer, typically by one order of magnitude, than the flares observed in
\src. It is their short duration and the fact that they occur in a system
with a main-sequence companion what makes the event in \src\ unique.

Fast and short-lived flares have been also reported for the newly suggested
class of supergiant X-ray binaries, known as supergiant fast X-ray
transients, SFXT \citep{negu06,smit06}. As in the classical supergiant
binaries these sporadic outbursts typically last for few hours
\citep{sgue06}. They differ in that SFXTs are transient sources and in the
complex structure of the outbursts, consisting of one or a few short
flares. However, at least two SFXTs have shown flares with a duration
comparable to that of \src, namely XTE J1739--302  \citep{saka02} and IGR
J17544--2619 \citep{gonz04}.

The physical explanation of the fast outbursts is not clear.  Accretion
from a smooth homogeneous stellar wind is ruled out. Instead, porous
(clumpy) winds are invoked. Fast outbursts would be due to the accretion of
large, dense clumps on the neutron star \citep{walt07,negu08}. 
Recently, \citet{sido07} has proposed the presence of a second wind
component in the form of an equatorial disk around the supergiant donor to
explain the outbursts seen in the SFXT IGR J11215--5952, the only SFXT that
exhibits periodic outbursts. These outbursts last for about 10 days and
exhibit flaring activity near the peak.  They are reminiscent of the
type I outbursts so common among Be/X-ray binaries and most likely have a
complete different origin to that of the fast, short ($\sim 1$ hr)
flares. Even \citet{sido07} suggest that the $\sim$1-hr duration flares
seen at the peak of the longer outburst in IGR J11215--5952 can be ascribed
to the clumpy nature of the wind in the equatorial disk.

The asymmetry of mass outflows in Be stars is well established
\citep{wate88}. The structure of the stellar winds in Be stars  contains
two components: at higher latitudes, mass is lost through the
high-velocity, low-density wind (typical of early-type stars); in the
equatorial regions a slower and denser wind operates and ultimately forms
the disk. The interaction between the neutron star and the equatorial disk
gives rise to enhanced accretion and increase of the X-ray flux.

The short-lived flares in \src\ can then be explained by the interaction
between the accreting compact object and a clumpy stellar wind. Whether
this stellar wind is the polar or equatorial component is not clear and
will probably depend on the geometry of the orbit. If the equatorial plane
and orbital plane are not coplanar and the orbit is relatively wide (so
that the compact object does not directly crosses the disk) then the
neutron star will be exposed to the polar wind. The determination of the
orbital parameters of \src\ may provide the key to understand the physical
mechanism behind the flares.

% Section on bursting pulsar--------------------------------------------------------

Given the short duration of the flares in \src\ it is tempting to compare
the \src\ properties with those of neutron star low-mass binaries, in
particular those of the "bursting pulsar" and the "Rapid Burster" since
these systems also show fast and short episodes of increase X-ray emission.
Instead of the term "flares" the word "burst" is used in this context. 
Both systems are transient, and the accretor in both is a neutron star.
Both show postburst "dips" followed by a relatively long recovery interval
preceding a further burst. This behavior can be understood in terms of
depletion of the reservoir followed by a fill-in time. Though the burst
duration is very long in the case of \src\ it is also interesting to note
that burst durations up to 700\,s have been detected from the Rapid Burster
from type-II bursts. The longer bursts of Rapid Burster (comparable to those
observed in \src) have a recovery time of $\sim$1hr, clearly not covered by
the relatively short (1hr) RXTE pointings. 

Type II bursts likely originate by spasmodic release of gravitational
potential energy, which is almost certainly due to some not yet understood
accretion disk instability. In support of this similarity of bursts
is the fact that the spectra of the burst and the persistent emission are
nearly identical, suggesting a similar (rescaled) mechanism for both
\citep{brig96,lewi96}.

% End of section on bursting pulsar--------------------------------------------------------

\section{Conclusions}

The available observational data indicate that \src\ is a member of the
class of massive X-ray binaries known as Be/X. The relatively short spin
period,  long orbital period, spectral components, transient emission and
strong H$\alpha$ emission are typical of this type of systems. However, the
short-lived flares make \src\ unique in its class.

The time-average energy spectrum  can be represented by an absorbed power
law ($\Gamma\approx0.7$), modified at high energy by a  cut-off ($E_{\rm
cut}\approx4.8$ keV, $E_{\rm fold}\approx9$ keV. An Fe line at $\sim$6.4
keV as also present. The out-of-flare power spectra show a QPO at $\sim$ 1
Hz, which is not present in the flare power spectra.

The out-of-flare pulses are consistent with an average profile. During the
flare (and three early cycles around cycle 35), the cycle can be decomposed
into the enhanced template of the out-of-flare pulse, plus a  main
``Gaussian'' component starting at phase 0.5 and drifting monotonically to
phase  0.6. Its width is about 0.15-0.3 in phase. During the post flare,
the curve is consistent with a de-enhanced template and a negative Gaussian
component of about the same width as the previous one and a phase around
0.5.

Although the actual triggering mechanism of the flares is not known, we
speculate that it can be associated to a non-homogeneous, i.e. clumpy
stellar wind.

\begin{acknowledgements}

This work has been  supported in part by the European Union Marie Curie
grant MTKD-CT-2006-039965.  This work has made use of NASA's Astrophysics
Data System Bibliographic Services and of the SIMBAD database, operated at
the CDS, Strasbourg, France. The ASM light curve was obtained from the
quick-look results provided by the ASM/RXTE team.

\end{acknowledgements}

\end{document}